\documentclass[aps,prb,twocolumn,showpacs]{revtex4}

\usepackage{graphicx}
\usepackage{dcolumn}

\begin{document}

\title{Short range repulsive interatomic interactions in energetic processes in solids}

\author{J.M. Pruneda}
\email[Corresponding Author:]{mpru02@esc.cam.ac.uk}
\author{Emilio Artacho}
\affiliation{Department of Earth Sciences, University of Cambridge
Downing Street, Cambridge, CB2 3EQ, United Kingdom}

\date{\today}

\begin{abstract}
The repulsive interaction between two atoms at short 
distances is studied in order to explore the range of validity 
of standard first-principles simulation techniques and improve 
the available short-range potentials for the description of 
energetic collision cascades in solids.  Pseudopotentials represent 
the weakest approximation, given their lack of explicit Pauli 
repulsion in the core-core interactions.  The energy (distance) 
scale realistically accessible is studied by comparison with 
all-electron reference calculations in some binary systems.  Reference 
calculations are performed with no approximations related to 
either core (frozen core, augmentation spheres) or basis set.  
This is important since the validity of such approximations, 
even in all-electron calculations, rely on the small core 
perturbation usual in low-energy studies.  The expected importance of 
semicore states is quantified.  We propose a scheme for 
improving the electronic screening given by pseudopotentials for 
very short distances.  The results of this study are applied to the 
assessment and improvement of existing repulsive empirical potentials.
\end{abstract}


\maketitle

\section{Introduction}

A high-energy ion (tens to hundreds of keV)
propagating through the system can approach other atoms to 
very close distances, typically between 1 \AA\ and 0.1 \AA.  
The strong repulsive potential determines the scattering process.  
Ion implantation, laser ablation and 
displacement cascades produced by $\alpha$-decay events in 
materials used for nuclear waste immobilization, are examples in which
these energetic collisions play an important role.  The energetic ion
loses energy due to inelastic collisions with the electrons and other atoms in
the material\cite{Bohr}.  The latter, called nuclear stopping, dominates for
relatively low energies ($\lesssim$ 1 keV/amu), while electronic stopping
dominates for higher energies.

Nuclear stopping processes represent an enormous challenge 
because the kinetic energies involved allow for 
displacements of the atoms in large regions, and the
energy dissipation takes place over long time scales 
(at least tens of picoseconds). Complicated structural 
configurations are obtained, with atoms breaking and making new bonds, 
and ionisation processes affecting the interactions between the particles.
The complexity in physical and chemical interactions, the 
large sizes, and the long times involved in the problem
demands high-efficiency first-principles calculations.

Many Density Functional Theory (DFT)\cite{DFT} implementations 
used by the condensed matter community
are based in the pseudopotential approximation, in which the core
electrons are considered as frozen in their atomic configuration,
and only the valence electrons are responsible of the chemical and
physical properties of the solid, resulting in a substantial
reduction of the degrees of freedom to be solved and hence in much
faster calculations than the equivalent all-electron.  
For the very challenging problems in this area, it would be highly 
desirable to asses the usefulness and the range of validity of such 
approximation.  It has to be remembered, however, that 
conventional all-electron approaches are not
much better suited for very short distances.
If augmented plane waves (APW) are used to solve numerically the 
all-electron problem in the core region, problems arise when the 
augmentation spheres overlap. The need for an accurate basis set 
for the core electrons might prevent the use of available gaussian 
basis sets, normally devised for unperturbed core electrons. 

Reference calculations are thus needed to test the accuracy of 
pseudopotential-based first principles calculations for 
describing electronic screening at short distances.  Experimental 
uncertainties in the determination of the potentials are usually of 
the order of 10\% or more.  We have chosen to follow the approach of Nordlund 
{\it et. al.}\cite{Nordlund}, and use as a reference calculations at 
the Hartree-Fock limit, where all these problems are eluded and the 
only remaining approximations are the absence of electronic correlation 
and of relativistic effects.
In this work we have studied the relative importance of each of 
the effects that could be relevant to the problem, and assessed 
the validity of the pseudopotential approximation.

\section{Methodology}
For high energies the interatomic separation in a head-on collision 
can be small, making the binary collision (BC) an appropriate 
approximation. In this case, the complexities of the repulsive 
interaction between atoms can be modelled by a simple two-body 
interatomic potential for short distances.  The interatomic potential 
$V(R)$ is defined as the difference between the total energy of the 
two atoms at a distance $R$, and the energy of the isolated atoms:
\begin{equation}
V(R)=E(R)-E(\infty)
\end{equation}

We use two descriptions to characterize this potential: $\delta V(R)$,
and the screening function $\Phi(R)$.  The former is defined as 
\begin{equation}
\delta V(R)= V(R)-\frac{Z_1 Z_2 e^2}{R}
\end{equation}
where $Z_1$ and $Z_2$ are the atomic numbers of the two atoms.
This can be used to characterize the potentials at 
$r\rightarrow 0$.  In the limit, we have the joint-atom solution, 
with just one atom with atomic number $Z_1+Z_2$.
Notice that $\delta V(0) = E^{atom}(Z_1+Z_2) - E^{atom}(Z_1) - E^{atom}(Z_2)$, 
and we can use all-electron {\it atomic} calculations (without the problems 
related to basis commented before) to obtain information of this limit of the
repulsive potential.

A different recasting of the interatomic potential that is also very 
useful\cite{Smith}, is obtained if we consider the effect of screening due to 
the electronic cloud around the nuclei:
\begin{equation}
V(R)=\frac{Z_1 Z_2 e^2}{R}\Phi(R)
\label{Phi}
\end{equation}
where $\Phi(R)$ is defined as the screening function.  For very short
distances, there is almost no screening of the internuclear repulsion 
because the probability of finding electrons in the interatomic region 
is very small, and $\Phi(0)\longrightarrow 1$.  
For long distances the potential goes to zero faster than the Coulomb
potential and $\Phi(\infty)\longrightarrow 0$ (for neutral atoms).  
Some popular parametrizations of this potential\cite{Torrens} are the 
ones due to Moli\`ere, Lenz-Jenson, Born-Mayer, and the one by Ziegler, 
Biersack and Littmark\cite{ZBL} (ZBL). We will 
compare our results with the latter, that takes the form:
\begin{equation}
\Phi(R) = \sum_{i=1}^4a_i e^{-b_ir}
\end{equation}
where the numerical coefficients $a_i$ and $b_i$ are fitted to a
variety of interatomic repulsive potentials for different pairs.
This gives a relatively good agreement
with experiment, specially for light elements.  Nevertheless, the
deviation with experimental ion scattering\cite{expr} data can be as 
large as 20\%.

The DFT energy $E(R)$ is obtained with the Siesta method\cite{siesta} 
using separable\cite{kb} norm-conserving Troullier-Martins 
pseudopotentials\cite{tm2}.  The valence wave functions are expanded 
in pseudoatomic numerical orbitals including multiple-$\zeta$ and 
polarization function.  Periodic boundary conditions
require the definition of a supercell large enough for the interaction
between replicas of the atoms to be sufficiently small (L$\sim$20\AA).  An
uniform real-space grid defined by an equivalent plane-wave cutoff
of 350 Ry was used for numerical integration.

In our simulations, scalar relativistic pseudopotentials are 
generated using different electronic configurations, including 
semicore states, partial core corrections, and different 
exchange-correlation functionals.   The use
of semicore shells, that is, core electrons considered as part of the
valence, has the benefit of increasing the accuracy as compared to
all-electron calculations, making the pseudopotential harder, and more
demanding computationally.  In table \ref{pseudos} the different 
configurations used for core/valence electrons are shown.  Partial 
core corrections\cite{PCC} (PCC) account at least partially for 
nonlinearity of the exchange-correlation functional due to the 
charge-density overlap between core and valence electrons.
Notice that including scalar 
relativistic effects, as well as the correlation,  represents, at 
least in some respects, an improvement to HF.  

\begin{table}[b]
\caption{ Electronic configurations considered for the
pseudopotential generation.}
\begin{tabular*}{8.0cm}{c@{\extracolsep{\fill}}ccc}
\hline
\hline
   & {core} & {\it semicore} & {valence} \\
\hline
C  &                        & 1s$^2$        & 2s$^2$ 2p$^2$         \\
O  & [He]                   &               & 2s$^2$ 2p$^4$         \\
Si & [He] 2s$^2$            & 2p$^6$        & 3s$^2$ 3p$^2$         \\
Ca & [Ne]                   & 3s$^2$ 3p$^6$ & 4s$^2$         \\
U  & [Xe]4f$^{14}$5d$^{10}$ & 6s$^2$ 6p$^6$ & 7s$^2$ 6d$^1$ 5f$^3$          \\
\hline
\hline
\end{tabular*}
\label{pseudos}
\end{table}

We compare the potentials obtained with Siesta with those obtained using 
the same numerical Hartree-Fock method as Nordlund 
{\it et. al.}\cite{Nordlund} and developed by Laaksonen 
{\it et. al.}\cite{2dhf}.  This HF approach, uses wave-functions of 
the general form:
\begin{equation}
\psi(\xi,\eta,\phi)=\frac{e^{im\phi}}{(2\pi)^{1/2}}u(\xi,\eta)
\end{equation}
and the Hartree-Fock equations are reduced to partial differential
equations for the function $u(\xi,\eta)$ in prolate spheroidal
coordinates.  The equations are then discretized using finite 
differences.

This numerical method is used as a reference in our electronic 
structure calculation.  The effect of electronic correlation was 
shown not to be relevant, at least for the light element potential 
studied by Nordlund {\it et. al.}\cite{Nordlund}.  We confirm this result on 
other systems. We also checked that 
relativistic effects not included in the Hartree-Fock method, 
are important but not relevant for the purpose of this work.  Thus,
albeit the computational cost of HF restricts its application to relatively
simple systems, it appears to be a good tool to gauge our approximation 
and test the accuracy achieved with other methods.

\section{Results}

\begin{figure}[t!]
\includegraphics[scale=0.48]{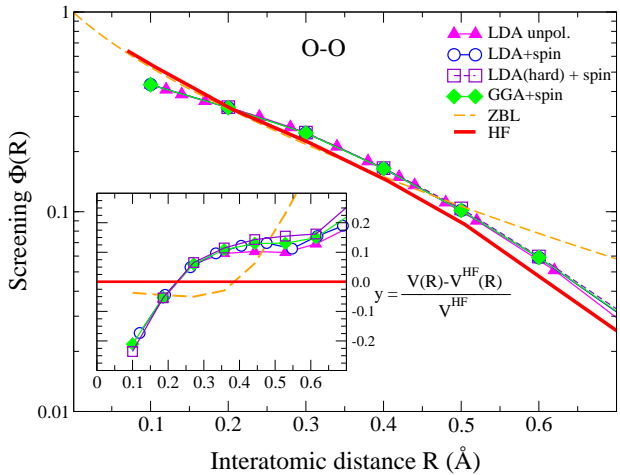}
\caption{(Color online) Screening function for O-O with different approximation 
for exchange-correlation (spin unpolarized LDA, spin polarized LDA 
and GGA, HF and ZBL). The reference configuration for the 
pseudopotential generation was 2s$^2$2p$^4$ with core radius 
$r_c=0.61$\AA,  except for the {\it hard} LDA, where $r_c=0.53$\AA.
The inset shows the ratio between the difference in 
computed potentials and the HF result.}
\label{O-O}
\end{figure}

The pairs considered in this work were C-C, O-O, Si-Si, Ca-O, and 
Ca-Ca with both HF and DFT, and U-O with only DFT.
The screening functions $\Phi(R)$ are shown in figures \ref{O-O} and
\ref{semicores}.  Deviations from HF become relevant for distances 
$R\ge0.7$\AA, when the attractive bonding interaction begins to be 
important.  It is in this region of bonding character in the interatomic 
potential where the different descriptions of exchange and correlation 
give different results. It is, however, not important for this study.
We used the Ceperley-Alder\cite{ca} parametrization of exchange-correlation
in the Local Density Approximation (LDA), and the Perdew, Burke and
Ernzerhof\cite{pbe} scheme for Gradient-corrected functional (GGA),
both with and without spin polarization.  We observed (Fig. \ref{O-O})
that any of these approximations for exchange and correlation
gives essentially the same repulsive potential.  
The agreement with HF is good in general, even for
short distances, showing that electronic correlation plays only a minor
role in the repulsive potential.  
Note that ZBL gives a good description of the repulsive HF potential for 
light elements (C, O).  For larger elements this agreement is not so good 
(see below).  

\subsection{Analysis of the different approximations}

\subsubsection{Effect of semicore states}

\begin{figure*}[t]
\includegraphics[scale=0.78]{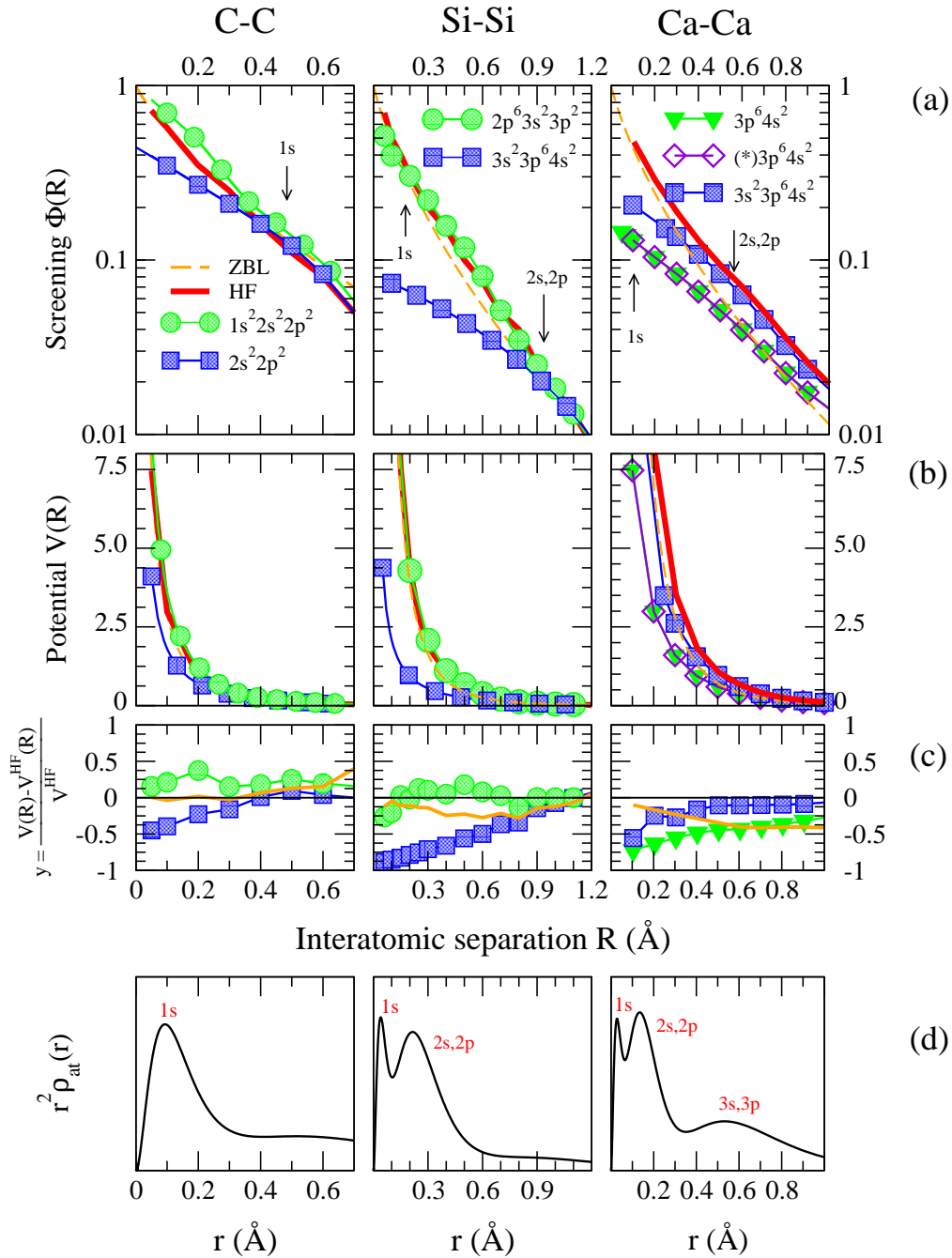}
\caption{(Color online). (a) Screening function for C-C, Si-Si and Ca-Ca, with different
electronic configurations for the pseudopotential. (b) Interatomic 
repulsive potential $V(R)$ in keV. 
(c) Deviation of the potential with respect to the HF limit.  
(d) Charge distribution, $r^2\rho(r)$, showing the core electronic shells.
 $(*)3p^64s^2$ differs from $3p^64s^2$ just in the number of $\zeta$'s used to
describe the 3p orbitals (double-$\zeta$, and single-$\zeta$ respectively).
The small arrows in (a) show the distance at which electronic clouds 
from the core start to overlap with each other (R=2$r_{90}$, see 
table \ref{ratomic}).}
\label{semicores}
\end{figure*}

In Fig. \ref{semicores} we analyze the influence of semicore states
in the screening.  When the core electrons of one atom
begin to overlap with the core electrons of the other atom, the
Pauli repulsion between them is not properly taken into account in the
pseudopotential approximation and consequently the result deviates from the
Hartree-Fock description.  The use of semicore states in the pseudopotential
improves this, with part of the core included as valence. 
This can clearly be seen in the top Fig.2a, comparing the HF results 
with those obtained with pseudopotentials.  In 2b, we observe the 
influence of these semicore states in the hardness/softness of the 
repulsive potential.  The absence of semicore electrons in 
the valence, makes $V(R)$ softer.  Fig.2c shows the deviation of the 
potential with respect to the HF value.  Although this result was 
expected, it is important to quantify it with respect to other effects, 
and to obtain an easy rule to assess the limit of validity of the 
pseudopotential approximation.

\begin{table*}[]
\caption{Radii (in \AA) containing 
90\% or 99\% of the total charge for each atomic orbital
for C, O, Si and Ca. Core electron 
eigenenergies ($\epsilon$ in eV) are also shown as computed with an 
all-electron atomic DFT, as well as experimental values for the 
corresponding binding energies\cite{webelements}.}
\begin{tabular*}{17cm}{c@{\extracolsep{\fill}}cccc|ccccc|ccccc|ccccc}
\hline
\hline
\multicolumn{5}{c|}{C}&\multicolumn{5}{c|}{O}&\multicolumn{5}{c|}{Si}&\multicolumn{5}{c}{Ca}\\
   shell &  $r_{90}$  & $r_{99}$ & $\epsilon$ & $E_{B}$  &
   shell &  $r_{90}$  & $r_{99}$ & $\epsilon$ & $E_{B}$  &
   shell &  $r_{90}$  & $r_{99}$ & $\epsilon$ & $E_{B}$  &
   shell &  $r_{90}$  & $r_{99}$ & $\epsilon$ & $E_{B}$  \\
\hline
   1s & 0.26 & 0.42 &  -273 &  -284 &
   1s & 0.19 & 0.31 &  -514 &  -543 &
   1s & 0.11 & 0.17 & -1774 & -1839 &
   1s & 0.07 & 0.12 & -3929 & -4038 \\
   2s & 1.34 & 2.06 &  -14  &       &
   2s & 0.98 & 1.51 &  -24  &  -41  &
   2s & 0.47 & 0.70 &  -160 &  -149 &
   2s & 0.30 & 0.44 &  -412 &  -438 \\
   2p & 1.64 & 2.71 &   -5  &       &
   2p & 1.17 & 1.96 &   -9  &       &
   2p & 0.48 & 0.77 &  -117 &  -100 &
   2p & 0.29 & 0.44 &  -326 &  -349 \\
      &      &      &       &       &
      &      &      &       &       &
   3s & 1.55 & 2.17 &   -28 &       &
   3s & 0.92 & 1.32 &   -47 &   -44 \\
      &      &      &       &       &
      &      &      &       &       &
      &      &      &       &       &
   3p & 1.04 & 1.55 &   -28 &   -25 \\
\hline
\hline
\end{tabular*}
\label{ratomic}
\end{table*}

\begin{figure}[b]
\vspace{-0.3cm}
\begin{minipage}[t]{0.49\linewidth}
\hspace{-3.0cm}
\begin{minipage}[]{0.23\linewidth}
\includegraphics[width=3.85cm]{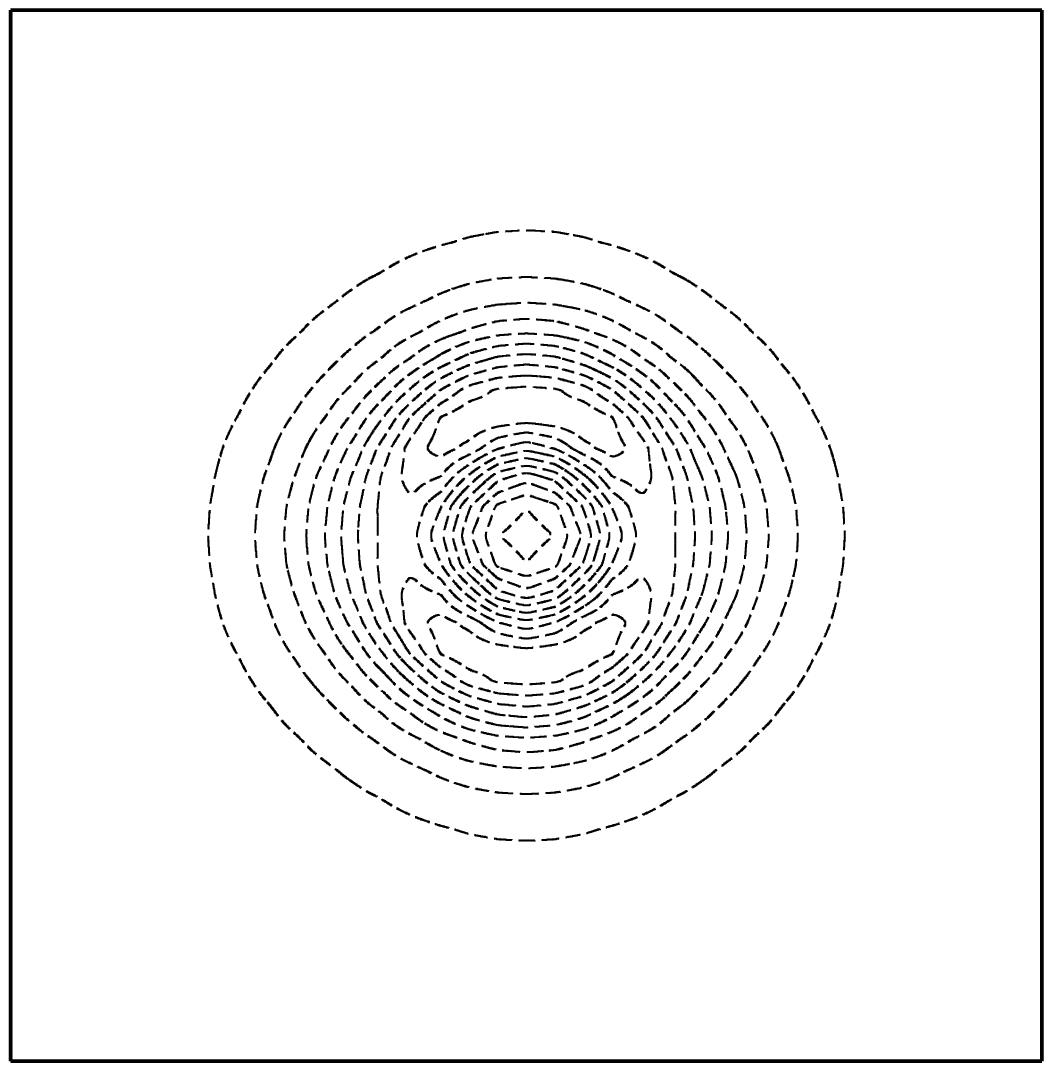}
\end{minipage}
\hspace{2.5cm}
\begin{minipage}[]{0.23\linewidth}
\includegraphics[width=3.85cm]{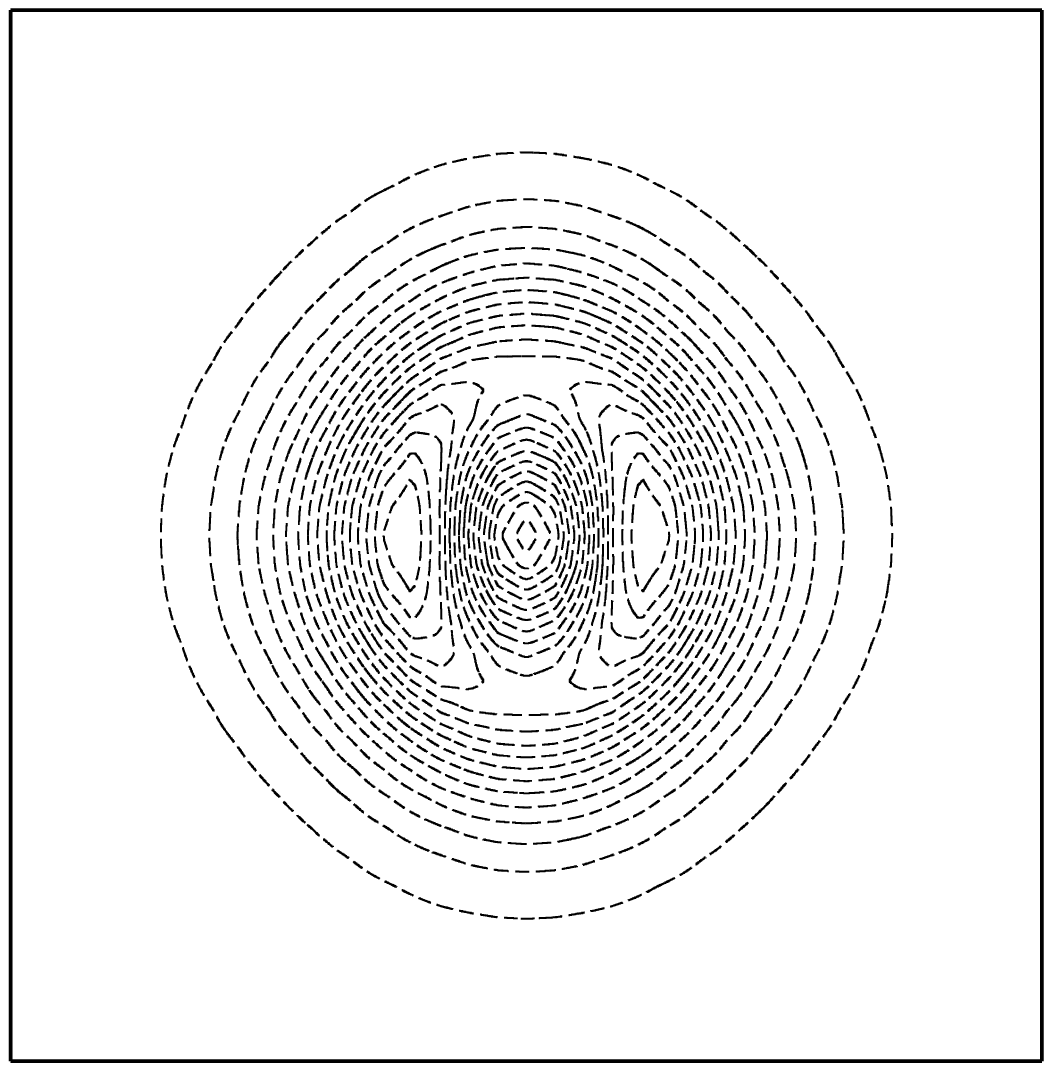}
\end{minipage}
\end{minipage}\\
\vspace{-0.2cm}
\begin{minipage}[b]{0.49\linewidth}
\hspace{-3.0cm}
\begin{minipage}[]{0.23\linewidth}
\includegraphics[width=3.85cm]{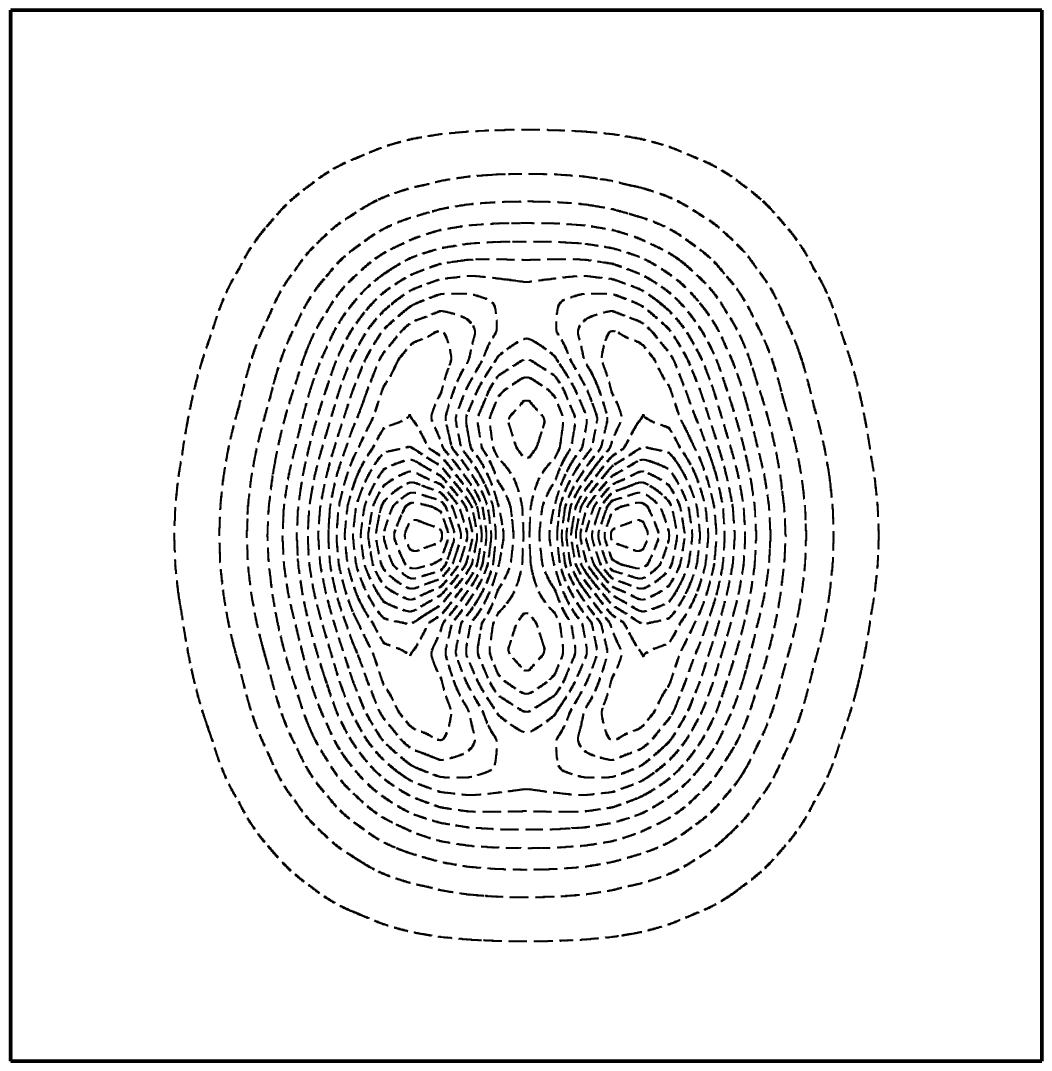}
\end{minipage}
\hspace{2.5cm}
\begin{minipage}[]{0.23\linewidth}
\includegraphics[width=3.85cm]{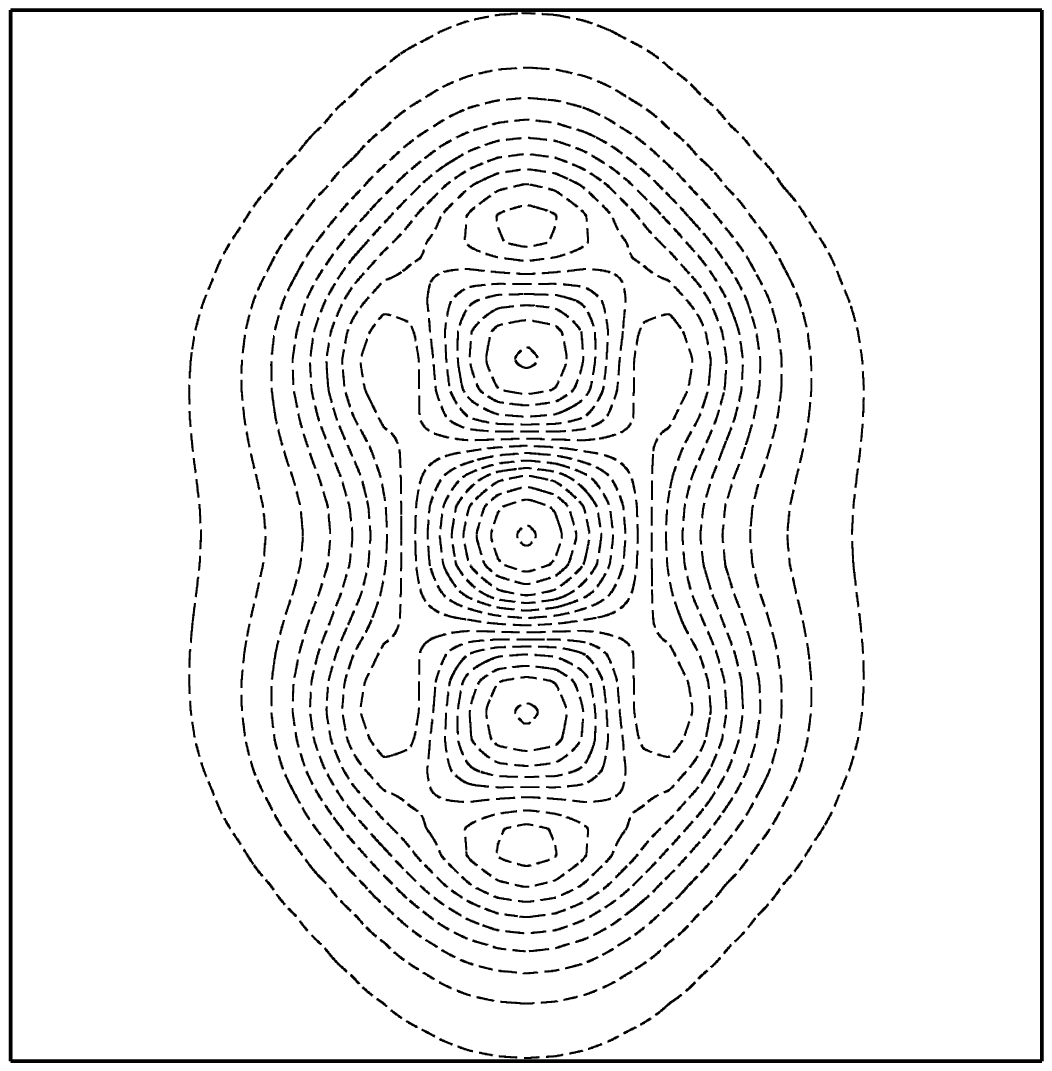}
\end{minipage}
\end{minipage}
\caption{Contour plot for the charge distribution at different 
interatomic distances for the Ca-Ca pair (0.1, 0.3, 0.5 and 1\AA,
going from left to right and up to down).}
\label{contour}
\end{figure}

When the interatomic distance is comparable to bonding distances
(R$\sim$1\AA), the valence electrons are spread around the atoms
and in the bonding region, screening the Coulomb interaction
between the nuclei.  As the distance between the nuclei is
reduced, the repulsive electronic interactions (Coulomb and Pauli) 
in the bonding
region increase, and the electrons are more unlikely to be in the
confined area between the nuclei.  This is shown in Fig. \ref{contour}, 
where the valence charge distribution around the
atoms is displaced from the interatomic center, to the surrounding
area as the distance decreases. In a simulation within the
pseudopotential approximation, the valence electrons have a
smaller participation in the nuclear screening, but the core
electrons are assumed to be fixed in their initial configuration,
and thus, their contribution to the screening is roughly the same.
For this reason $\Phi(R)$ is always smaller than the expected
screening function with an {\it all electron} approach.

Core electrons deep in the inner shells (see Fig.2d), are more 
strongly bonded to the nuclei and their participation in the collision 
would only be relevant for very high collision energies when the interatomic 
distance becomes really small.  
For that energy scale, the electronic stopping power is more important 
than the nuclear stopping power. Hence, an accurate description of the 
repulsive potential for shorter (higher) distances (energies), is beyond 
the scope of this study.

Only the outer core-shell electrons are included in the semicore.  
How localized these electrons are can be roughly estimated using 
the spread of the corresponding atomic Kohn-Sham orbitals in a 
DFT all-electron calculation.  In table \ref{ratomic} we 
show for each atomic shell, the radii $r_{90}$ and $r_{99}$,
in which 90\% or 99\% of the atomic 
wavefunction's norm is localized.  Two shells corresponding 
to different atoms will likely interact, if the interatomic distance 
is smaller than sum of the radii at which the charge is more localized.
According to this, and considering $r_{90}$ as a reference, in a Si-Si 
collision, we can say that 2s electrons will interact with 2s or 2p 
valence electrons in the other atom for distances R$\lesssim$0.95\AA, 
which is an appreciable distance (see arrows in Fig.2a).  
So, if we want to have a good 
description of the collision for shorter ranges, we would need to 
include 2s electrons into the semicore.  With that, we would ensure 
a reasonable description of the interactions up to distances of the 
order of 0.2\AA, when the 1s electrons start to play a role.  This 
would explain why the pseudopotential without any semicore state 
gives such a soft repulsive potential.

The same argument may be used for 1s electrons in a C-C
collision for distances smaller than 0.4\AA, and for the 2s-2p electrons
in a Ca-Ca collision below $\sim$0.6\AA.  In the Ca-Ca case, both the 3s 
and 3p shells have to be included in the semicore for reasonable screening 
down to 2$r_{90}$, in accordance to the model.  In Si-Si, however, the 2p 
electrons seem to be enough, and the 2s apparently do not affect the 
screening for distances smaller than $\sim$1\AA.

\subsubsection{Effect of the core radius}
In the pseudopotential approximation, it is assumed that the core electrons 
of different atoms do not overlap.  For the interatomic distances of 
interest in this work, this is not the case.  Although the core radii of the 
pseudopotentials ($r_c$) play an important role in the description of ground
state properties of real materials, the effects for the repulsive potential 
at short distances is not as dramatic.  Changes of up to 40\% in $r_c$ 
give differences in the repulsive potentials smaller than 2\% when the 
interatomic distance is smaller than 2$r_c$, whereas differences can be 
as large as 20\% when the distance is close to 2$r_c$.

\subsubsection{Effect of charge state, correlation, and 
relativistic corrections}

The charge state of the shooting ion can play an important role in the 
description of the electronic energy loss for intermediate energies.  
For the short range interatomic repulsion, however, these effects are 
not so relevant: since electrons move out of the internuclear region
for short interatomic distances (Fig. \ref{contour}), there will be 
no important effect on the repulsive potential if the ionization 
involved not too deep electrons.

\begin{figure}[b!]
\includegraphics[scale=0.48]{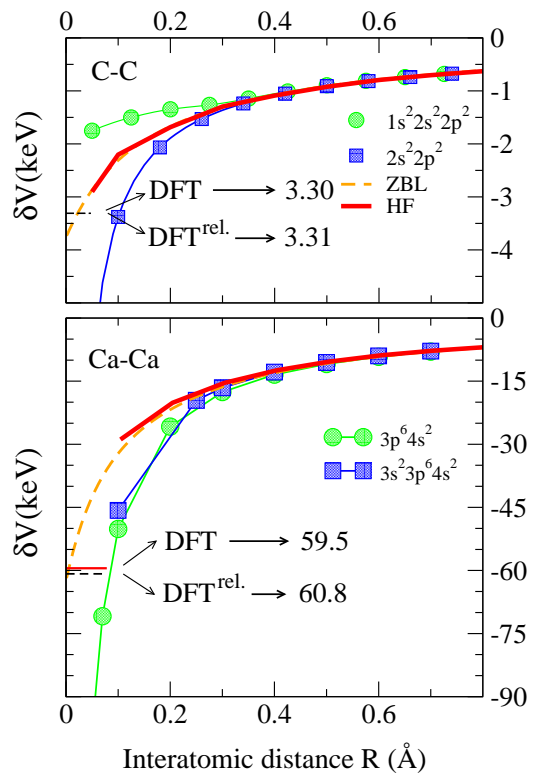}
\caption{(Color online) Non-coulombic interaction, $\delta V(R)$, 
for C-C (top) and Ca-Ca (bottom), and comparison with 
relativistic effects. The solid ticks at $\sim$3.3 keV (for C-C) and 
$\sim$60 keV (Ca-Ca) represent the atomic energy of Mg and Zr (atomic 
numbers twice as big as C and Ca) with an {\it all-electron} atomic DFT.  
The dashed tickmark represents the atomic energy with relativistic 
corrections (for C if falls over the solid lines in the energy scale used).
}
\label{relativistic}
\end{figure}
Correlation and relativistic corrections are known to be relevant 
for very short distances or very heavy nuclei.  
In the pseudopotential method, scalar relativistic corrections are included,
but not in the HF approach that we use as a reference.
With an all-electron DFT-based {\it atomic} calculation which includes 
relativistic effects (scalar and spin-orbit), we use $\delta V(0)$ 
to estimate an upper limit of the relativistic corrections.  As an example 
a limit of $\sim$1.3 keV is obtained in a Ca-Ca collision.  Although this is 
a considerable correction, the energy scales 
we are interested in (see Fig. \ref{relativistic}), are of the order of tens 
of keV. Furthermore, the relativistic effects are expected to be smaller 
for distances $\gtrsim$0.4\AA.  The difference between DFT and HF 
(both relativistic or both non-relativistic) is of the order 
of $\sim$5 eV and gives an idea of the 
magnitude of correlation effects.  It is evident from the figure that the 
approximations involved in the pseudopotential approach are much more 
relevant than correlation or relativistic effects.

\subsection{Practical realization: Heavy ions}
We can now try to obtain some qualitative information about the 
interatomic potential for heavy ions.  The use of pseudopotentials 
for equilibrium (low energy) situations of lanthanide and actinide 
compounds has been shown to be accurate\cite{Pickard,Louie} 
when the effects coming from overlap 
between core and valence are properly treated (either with the 
inclusion of semicore states, or with partial core correction), 
even if only scalar relativistic corrections are considered.  
Relativistic spin-orbit terms can also be important in 
these elements, and the results obtained within this method should be 
taken with caution.

\begin{table*}[b]
\caption{ Estimated radii (\AA) for the localization of electronic orbitals 
in U, eigenenergies (eV) as obtained in an atomic DFT calculation, and 
electron binding energies (eV)\cite{webelements}.}
\begin{tabular*}{18cm}{c@{\extracolsep{\fill}}|cccc||c|cccc||c|cccc||c|cccc}
\hline
\hline
   shell   &  $r_{90}$  & $r_{99}$ & $\epsilon$ & $E_{B}$  &
   shell   &  $r_{90}$  & $r_{99}$ & $\epsilon$ & $E_{B}$  &
   shell   &  $r_{90}$  & $r_{99}$ & $\epsilon$ & $E_{B}$  &
   shell   &  $r_{90}$  & $r_{99}$ & $\epsilon$ & $E_{B}$  \\
\hline
1s         & 0.01 & 0.02 & -114939 & -115606 &
3p$_{3/2}$ & 0.12 & 0.17 &   -4222 & -4303 &
4d$_{3/2}$ & 0.28 & 0.38 &    -762 & -778 &
5p$_{3/2}$ & 0.59 & 0.79 &    -205 & -192 \\
2s         & 0.05 & 0.07 &  -21496 & -21757 &
3d$_{3/2}$ & 0.11 & 0.16 &   -3674 & -3728 &
4d$_{5/2}$ & 0.28 & 0.39 &    -720 & -736 &
5d$_{3/2}$ & 0.67 & 0.93 &    -116 & -103 \\
2p$_{1/2}$ & 0.04 & 0.06 &  -20731 & -20948 &
3d$_{5/2}$ & 0.11 & 0.16 &   -3496 & -3552 &
4f$_{5/2}$ & 0.29 & 0.42 &    -391 & -388 &
5d$_{5/2}$ & 0.69 & 0.97 &    -108 & -94 \\
2p$_{3/2}$ & 0.05 & 0.07 &  -16961 & -17166 &
4s         & 0.24 & 0.33 &   -1395 & -1439 &
4f$_{7/2}$ & 0.29 & 0.43 &    -380 & -377 &
6s         & 1.11 & 1.51 &     -60 & -44 \\
3s         & 0.11 & 0.15 &   -5440 & -5548 &
4p$_{1/2}$ & 0.24 & 0.33 &   -1238 & -1271 &
5s         & 0.50 & 0.67 &    -320 & -321 &
6p$_{1/2}$ & 1.26 & 1.74 &     -43 & -27 \\
3p$_{1/2}$ & 0.11 & 0.15 &   -5092 & -5182 &
4p$_{3/2}$ & 0.27 & 0.37 &   -1011 & -1043 &
5p$_{1/2}$ & 0.52 & 0.71 &    -259 & -257 &
6p$_{3/2}$ & 1.42 & 1.99 &     -34 & \\
\hline
\hline
\end{tabular*}
\label{BohrU}
\end{table*}

\begin{figure}[t]
\includegraphics[scale=0.38]{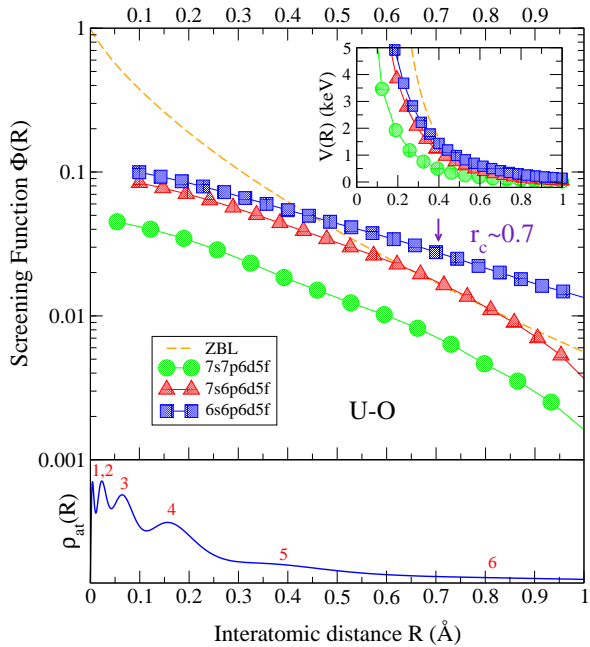}
\caption{(Color online) Screening function, $\Phi(R)$, and repulsive 
interatomic potential, $V(R)$ in the inset, for the U-O pair.  The arrow 
shows the radius at which the n=5 shell of U overlaps with the n=1 shell 
of O. Bottom figure represents the electronic charge distribution 
for U, as computed with an atomic all electron DFT method.}
\label{U-O}
\end{figure}

We have computed the uranium-oxygen interatomic 
potential $V(R)$ with different pseudopotentials, and compared 
it to ZBL (see Fig. \ref{U-O}).  The electronic configuration is 
shown in table \ref{pseudos}, and the localization radii $r_{90}$ 
and $r_{99}$, in table \ref{BohrU}.  Using the same arguments as before,
we could expect that using a semicore with 6s and 6p electrons in the 
valence, the U-O interatomic potential obtained would be appropriate up to 
distances of the order of 0.7\AA.  Note that the ZBL potential would give 
a lower screening than expected (at least for distances larger than 0.7\AA), 
giving a harder description of the binary collision.

\begin{figure*}[b]
\includegraphics[scale=0.78]{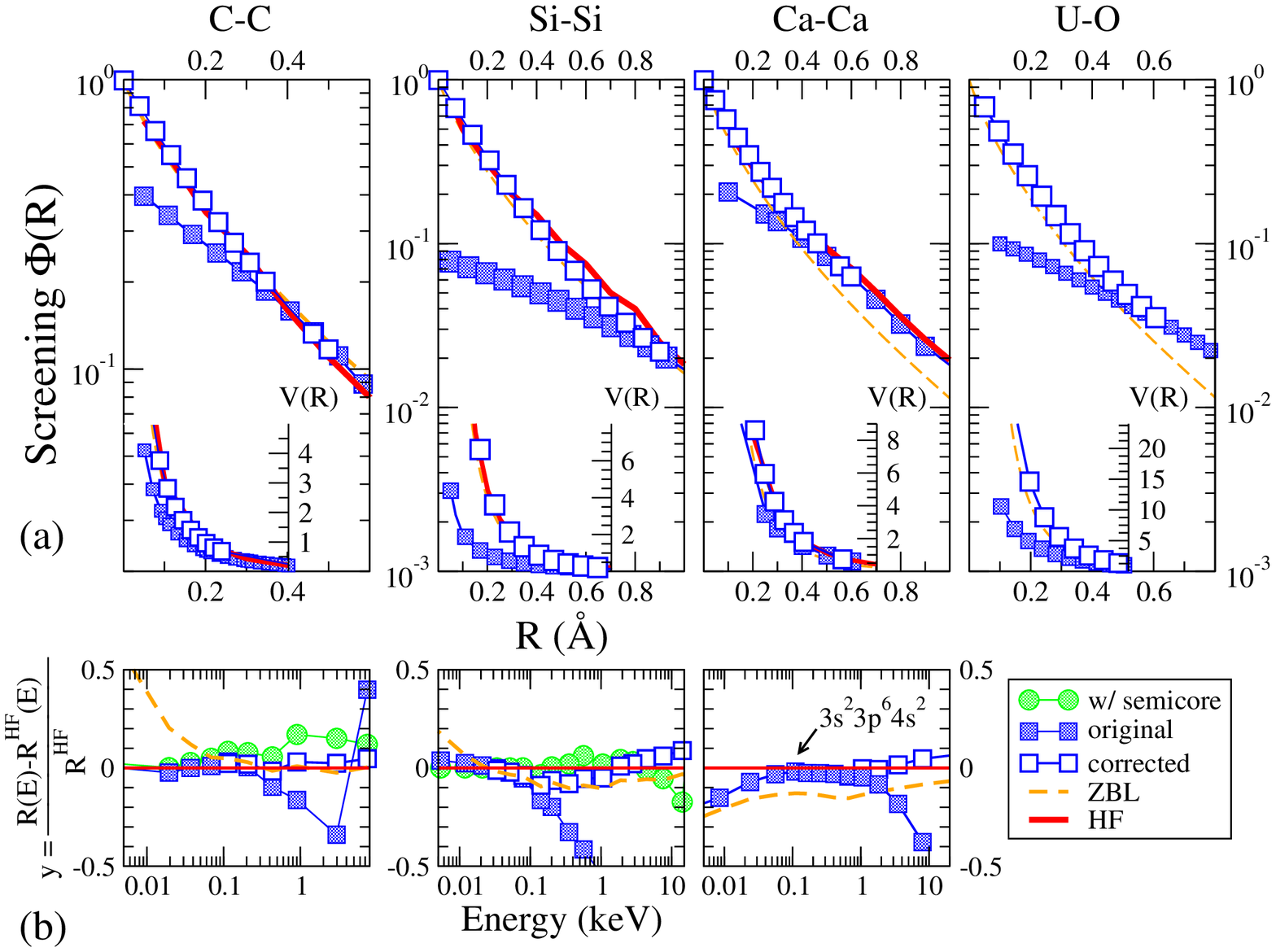}
\caption{(Color online). (a) Interatomic screening function for different pairs with the 
correction scheme proposed as compared with previous results.
(b) Relative deviation with respect to HF in the internuclear distance 
corresponding to each energy.  We compare the results obtained for: 
(1) the pseudopotentials without semicore (original); 
(2) with the correction proposed in this work (open square symbols); 
(3) the pseudopotential with semicore states; 
and (4) the ZBL parametrization.  
Note that considerable deviations in (1) appear for energies of hundreds 
of eV, and this is considerably improved with (2).  ZBL gives 
good description of the high energy regime, but fails for energies 
smaller than $\sim$0.1 keV.  }
\label{corrected}
\end{figure*}

\subsection{Correction of the screening function}
A systematic feature of the pseudopotential calculations is that 
in the limit of small distances, the screening functions obtained do not 
approach unity.  This is due to the frozen core approximation and the 
fact that the inner shells of the electronic cloud are not contributing 
to the screening.  We propose a method improve $\Phi$ by adding a 
continuous function that gives the proper limit ($\Phi(0)\rightarrow1$) 
for values below the 
radius $r_0$ at which the core states begin to overlap.  By inspection of 
figures \ref{semicores} and \ref{U-O} we could use the 
simplest function $\Phi(R)=e^{\alpha R^2+\beta R}$, with 
parameters $\alpha$ and $\beta$ fixed by the conditions over
$\Phi(r_0)$ and $\frac{d\Phi}{dR}(r_0)$, giving 
continuity of the screening and its derivative (and thus, the interatomic 
force) at $r_0$.  Notice that the matching radii would depend on the 
particular configuration used for the generation of the pseudopotential. 
We propose to use the sums of the $r_{90}$ for the relevant core states of
both atoms considered: $r_0=r_{90}^{atom1}+r_{90}^{atom2}$.
In figure \ref{corrected} we compare the previous 
screening functions with the ones corrected with this scheme.  
This will allow realistic 
molecular dynamic simulations with pseudopotentials, even if the atoms 
get extremely close in a high energy collision.  Numerical fits of these 
potentials have already been used to do semiempirical simulations of collision
cascades for studies of radiation damage in CaTiO$_3$\cite{Kostya}.

\section{Conclusions}

The validity of the pseudopotential approximation has been checked at close
distances as found in high energetic collisions in materials.
We have compared the interatomic repulsive
potential computed with an all-electron Hartree-Fock method, with
that obtained with pseudopotentials. In this HF reference method, no 
approximation is assumed with respect to the core shells, or the 
basis used to represent the electronic states.  The validity of 
the HF approximation itself has been assessed in order to use it 
as a reference for the pseudopotentials. The effect of different
approaches to treat the electronic correlation was previously
shown\cite{Nordlund} not to give substantial differences in the
screening function, a result confirmed in this study.  The relativistic 
corrections, though important, can
be considered small for the energy scales of interest in this work.

It has been shown that as the nuclear distance decreases, the
electrons are expelled from the internuclear region to the
surrounding area, and the screening is completely dominated by the
deeper electrons, bound at short length scales.  
If the core is frozen at length scales longer or comparable to 
the internuclear distance, the
screening is larger than expected and the pseudopotential
description fails. We have shown this effect to be the 
most important one in the problem, actually the only one of
relevance at the energy scales involved.  

We have quantified to what extent the use of deeper
electrons as part of the valence configuration is essential 
to describe the repulsive potential for relatively
short internuclear distances. The localization of the core
electrons, $r_{90}$, has been found as the most appropriate criterion 
used to decide which states have to be included for a particular 
range of collision energies.  The pseudopotential description would 
then be valid until the more external core shells not included in 
the calculation start to overlap.
Finally, we have proposed a scheme to
correct the screening given in the pseudopotential approximation, that
recovers the right trend for very short distances, and improves
the use of pseudopotentials for the description of energetic collisions with
first principles simulations.

\section{Acknowledgments}
This work was supported by British Nuclear Fuels (BNFL) and NERC.
We would like to thank M. Dove, K. Trachenko for helpful discussions and
sharing their experience with MD methods in radiation damage simulations.

\end{document}